\documentclass[showpacs,preprintnumbers,amsmath,amssymb,12pt,floatfix,epsfig]{revtex4}
\usepackage{graphicx}
\begin{document}
\draft
%234567890123456789012345678901234567890123456789012345678901234567890
\title{y-scaling in Quasielastic Electron Scattering from Nuclei}
\author{K. S. Kim$^{1)}$\thanks{kyungsik@hau.ac.kr}
and L. E. Wright$^{2)}$\thanks{wright@ohiou.edu}}
\address{1)School of Liberal Arts and Science, Korea Aerospace University,
Koyang 200-1, Korea  \\
2)Institute of Nuclear and Particle Physics, Ohio University,
Athens, OH 45701}

\begin{abstract}
A relativistic single particle model is used to calculate the
inclusive $(e,e')$ reaction from $A=$12, 40, 56, 197, and 208
nuclei in the quasielastic region. We have shown that this model
provides a very good description of the available experimental
cross sections when they are dominated by the quasielastic process.
In this paper we use this model to investigate the dependence  of
$y$-scaling on electron kinematics, particularly the electron
scattering angle, for a range  of squared four momentum transfer
$0.20-0.80$ (GeV/c)$^2$. In this kinematic domain, Coulomb
distortion of the electron does not significantly affect scaling,
but final state interactions of the knocked out nucleon do affect
scaling particularly when the nucleons have lower energies. In general, we find
that scaling works for this reaction, but at lower values of the four
momentum transfer, the scaling function does have some dependence on
the electron scattering angle. We also
consider a modification of y-scaling to include small binding
energy effects as a function of Z and A and show that there is
some improvement in scaling.
\end{abstract}
\narrowtext
\maketitle

%234567890123456789012345678901234567890123456789012345678901234567890
Medium and high energy electron scattering has long been
acknowledged as a useful tool to study nuclear structure and
nuclear properties, especially in the quasielastic region where
the process of knocking out nucleons is dominant. Many experiments
\cite{mezi,dead,bates,sacl} have been performed on medium  and
heavy nuclei at incident electron energies less than 1 GeV where
contributions of inelastic processes can be avoided. There are
also a number of theoretical works
\cite{jin1, bouc, chinn, trai, donn1, kim1, kim3, kim5} which have been
compared to the measured responses.

As we have noted before \cite{kim1,kim3,kim5}, the Fermi gas model
in the impulse approximation  roughly describes the inclusive
$(e,e')$ cross sections, but fails to provide a good  description
of the structure functions. A good theoretical description of
quasielastic scattering requires two ingredients before one can
compare experimental $(e,e')$ data from light to heavy nuclei to
theory. One of them is a model for the nuclear transition current
and the other is some provision for the inclusion of  electron
Coulomb distortion effects for medium and heavy nuclei.   In the
early 1990's, the Ohio University group \cite{jin1,jin2,zhang}
treated the electron Coulomb distortion exactly for the reactions
$(e,e')$ and $(e,e'p)$ in the quasielastic region using partial
wave expansions of the electron wave functions in the distorted
wave Born approximation (DWBA).  The Madrid group \cite{udias}
subsequently reported a similar analysis for the exclusive
$(e,e'p)$ reaction.  However, the DWBA calculations do not allow a
separation of the cross section into a longitudinal part and a
transverse part and are numerically challenging, and computational
time increases rapidly with higher incident electron energies. In
order to avoid these difficulties, Kim and Wright
\cite{kim1,kim3,kim2} developed an approximate treatment of the
electron Coulomb distortion which does allow the separation of the
cross section into a longitudinal part and a  transverse part.

In addition, we found a model which provides a very good
description of quasielastic scattering processes from nuclei
 for both the inclusive and exclusive cases. It is a
relativistic single particle model which requires the wave
functions of bound and continuum nucleons and a transition current
operator. The bound state wave functions are obtained from solving
the Dirac equation in the presence of the strong vector and scalar
potentials\cite{horo} and for the inclusive $(e,e')$ reaction
where the knocked out nucleons are not observed, the continuum
wave functions are solutions to a real potential so as not to lose
any flux. At low energies, this potential is just the same as the
bound nucleon potential and thereby guarantees current
conservation and gauge invariance. However, it is known from
elastic proton scattering that the continuum potential becomes
weaker with increasing proton energy. Therefore, for higher energy
processes we use a nucleon potential whose strength has been
fitted to proton elastic scattering \cite{clark,hung}. In a recent
paper \cite{kim6}, we found excellent agreement between our model
with the higher energy experimental data from SLAC \cite{day1} for
the quasielastic $(e,e')$ scattering on $^{12}$C, $^{56}$Fe, and
$^{197}$Au at the squared four momentum transfer of approximately
0.20 - 0.30 (GeV/c)$^2$ by using the energy-dependent real
potentials (which are weaker than the bound state potentials) for
the outgoing nucleons. Note that these calculations do not
conserve nucleon transition current, but we calculate all four
components of the transition current in order to minimize errors
due to lack of current conservation.  Note that the Madrid group
\cite{udias} has used a very similar relativistic model for
$(e,e'p)$ reactions.  The excellent agreement of this model with
experimental data allow us to use it as a tool for investigating
$y$-scaling of the quasielastic components of the inclusive
$(e,e')$ even in kinematic regions where inelastic contributions
have large contributions.

Since the pioneering work by West \cite{west}, there have been
many experimental \cite{sacl,day2} and theoretical
\cite{pace,ciof1,ciof2,ciof3,donn2,madrid} investigations of
$y$-scaling from nuclei.  In this paper we propose to use our
relativistic mean field single particle model with the inclusion
of Coulomb corrections to investigate the approach to $y$-scaling
at intermediate values of the four momentum transfer. Using
$y$-scaling, the measured cross section for the inclusive $(e,e')$
reaction can be written as a product of the electron-nucleon cross
section times a function $F$ which is related to the momentum
distribution of nucleons in the nucleus and is a function of
momentum transfer $q$ and energy transfer $\omega$. For the case
of large momentum transfer, the function $F$ should depend only on
a single variable $y$ which is a function of $\omega$  and $q$
\cite{ciof2}.  Scaling is expected to be valid for the very large
momentum transfer region, but it may be broken by final state
interactions in the quasielastic region and/or electron Coulomb
distortion effects.  A SLAC experiment \cite{day1} was performed
at the squared four momentum transfer $Q^2$ of 0.23 - 2.52
(GeV/c)$^2$ and the data at $y<0$ exhibited scaling at large
$Q^2$.   Recently, there have been additional experimental data
from JLAB \cite{arring} at 4.045 GeV, but scaling of the
quasielastic process cannot be demonstrated since the energy
transfer is sufficiently high that pion production is a
significant contribution to the cross section.  Note that both of
these experiments have been carried out at relatively forward
electron scattering angles. Furthermore, we cannot compare our
model predictions to these data since we have not yet included
inelastic processes (meson production, etc.)  in our model.
However,  as noted above we can use our model to investigate the
scaling of  the quasielastic contributions to the cross section.

Using non-relativistic models, the authors in Ref. \cite{ciof3}
analyzed $y$-scaling of the quasielastic electron scattering in
few-body system, complex nuclei, and nuclear matter. Within the
framework of the plane wave impulse approximation (PWIA), they
investigated the effects of the final state interaction, the
binding correction, and the nucleon-nucleon correlations. They
pointed out that the relation between the scaling function and the
momentum distribution does not exist at finite momentum transfer
because of the final state interaction and the binding correction.
More recently, the Madrid group in collaboration with Donnelly
\cite{anto, madrid,  cabal} have investigated scaling using a
semi-relativistic model and note that the strong scalar and vector
potentials in the final state of relativistic models result in a
breakdown of scaling and result in different scaling functions for
longitudinal and transverse responses.

In this paper, we initially calculate  $y$-scaling at the squared
four momentum $Q^2$ of approximately 0.2 - 0.3 (GeV/c)$^2$
comparing with the experimental data measured at SLAC \cite{day1}
for $^{12}$C, $^{56}$Fe, and $^{197}$Au, and Bates \cite{bates}
for $^{40}$Ca, where the quasielastic contribution is
kinematically isolated from pion production. We then investigate
the effects of final state interactions in our relativistic model
and electron Coulomb distortion on scaling.   Since we are
considering cases with large outgoing nucleon energies, we do use
an energy dependent final state interaction.  Finally, we
introduce a new $y$-scaling variable in order to solve the
non-scaling problem in the presence of the final state interaction
from different target nuclei in the same four momentum transfer
range.

In the plane wave Born approximation (PWBA), where the electron
wave functions are described by the Dirac plane waves, the cross
section for the inclusive quasielastic $(e,e')$ reaction is
written as
\begin{equation}
{\frac {d^{2}{\sigma}} {d{\omega}d{\Omega}_{e}}}
={\sigma}_{M} \left\{{\frac {q^{4}_{\mu}}{q^{4}}} S_{L}(q, \omega)
+({\tan^{2} {\frac {{\theta}_{e}}2}}-{\frac {q_{\mu}^{2}}{2q^{2}}})
S_{T}(q, \omega) \right \},
\label{cr}
\end{equation}
where ${q_{\mu}}^2={\omega}^2-{\bf q}^2 =-Q^2$ is the squared four
momentum  transfer, $\sigma_{M}=(\alpha/2E)^2
[{\cos}^2(\theta_e/2)/{\sin}^4(\theta_e/2)]$ is the Mott cross
section, and the longitudinal and transverse structure functions
which depend on the three momentum transfer $q$ and the energy
transfer $\omega$ are $S_{L}$ and $S_{T}$. Explicitly, the
structure functions for a given bound state with angular momentum
$j_{b}$ are given by
\begin{eqnarray}
S_{L}(q,{\omega})&=&\sum_{{\mu}_{b}s_{p}}{\frac {{\rho}_{p}}
{2(2j_{b}+1)}} \int {\mid}N_{0}{\mid}^{2}d{\Omega}_{p} \\
S_{T}(q,{\omega})&=&\sum_{{\mu}_{b}s_{p}}{\frac {{\rho}_{p}}
{2(2j_{b}+1)}} \int
({\mid}N_{x}{\mid}^{2}+{\mid}N_{y}{\mid}^{2})d{\Omega}_{p}
\end{eqnarray}
with the outgoing nucleon density of states ${\rho}_{p}={\frac
{pE_{p}} {(2\pi)^{3}}}$.
The ${\hat {\bf z}}$-axis is taken to be along the momentum transfer
${\bf q}$ and
 the z-components of the
angular momentum of the bound and continuum state nucleons are ${\mu}_{b}$ and
$s_{p}$, respectively.
The Fourier transform of the nuclear current $J^{\mu}({\bf r})$ is simply
given by
\begin{equation}
N^{\mu}=\int J^{\mu}({\bf r})e^{i{\bf q}{\cdot}{\bf r}}d^{3}r ,
\end{equation}
where $J^{\mu}({\bf r})$ denotes the nucleon transition current.
The continuity equation could be used to eliminate the
$z$-component ($N_{z}$) via the equation $N_{z}=-{\frac {\omega}
{q}}N_{0}$ if the current is conserved, but since we use an energy
dependent final state interaction \cite{kim6} we violate current
conservation and to minimize errors calculate $N_{z}$ directly.
The nucleon transition current in the relativistic single particle
model is given by
\begin{equation}
J^{\mu}({\bf r})=e{\bar{\psi}}_{p}({\bf r}){\hat {\bf J}}^{\mu}
{\psi}_{b}({\bf r}) \;,
\end{equation}
where ${\hat {\bf J}}^{\mu}$ is a free nucleon current operator, and
$\psi_{p}$ and $\psi_{b}$ are the wave functions of the knocked
out nucleon and the bound state, respectively.
For a free nucleon, the operator comprises the Dirac contribution and
the contribution of an anomalous magnetic moment $\mu_{T}$ given by
${\hat {\bf J}}^{\mu}=F_{1}(q_{\mu}^2){\gamma}^{\mu}+
F_{2}(q_{\mu}^2){\frac {i{\mu}_{T}} {2M}}{\sigma}^{\mu\nu}q_{\nu}$.
The form factors $F_{1}$ and $F_{2}$  are related to the electric and
magnetic Sachs form factors given by $G_{E}=F_{1}+{\frac {{\mu}_{T}
q_{\mu}^{2}} {4M^{2}}}F_{2}$ and $G_{M}=F_{1}+{\mu}_{T}F_{2}$ which
are assumed to take the following standard form:
\begin{equation}
G_{E}={\frac {1} {(1- {\frac {q^{2}_{\mu}}{\Lambda^2})^{2}}}}
={\frac {G_{M}} {({\mu}_{T}+1)}} \;,
\end{equation}
where the standard value for $\Lambda^2$ is 0.71 (GeV/c)$^2$.

The $y$-scaling function is defined as the ratio of the measured cross
section to the off-shell electron-nucleon cross section as follows:
\begin{equation}
F(y)={\frac {d^{2}{\sigma}} {d{\omega}d{\Omega}_{e}}}
( Z {\sigma}_{ep} + N {\sigma}_{en} )^{-1}
{\frac {q} {[M^2 + ( y+q)^2]^{1/2}}},   \label{fy}
\end{equation}
where ${\sigma}_{ep}$ (${\sigma}_{en}$) denotes the off-shell
electron-proton(neutron) cross section $\sigma_{cc1}$ defined by
Ref. \cite{defo}. In Eq. (\ref{fy}), $Z$ and $N$ are the number of
protons and neutrons, and $M$ is the mass of nucleon. The scaling
variable $y$ \cite{pace1} is given by
\begin{equation}
\omega + M_{A} = (M^2 + q^2 + y^2 + 2yq) ^{1/2}
+ ( {M_{A-1}}^2 + y^2 )^{1/2},
\end{equation}
where $M_{A}$ is the mass of the target nucleus and $M_{A-1}$ is
the mass of the ground state of the $A-1$ nucleus. The point $y=0$
corresponds approximately to the peak of the quasielastic
scattering and $y<0$ ($y>0$) corresponds to the small (large)
$\omega$ region.

\begin{figure}
\includegraphics[width=1\linewidth]{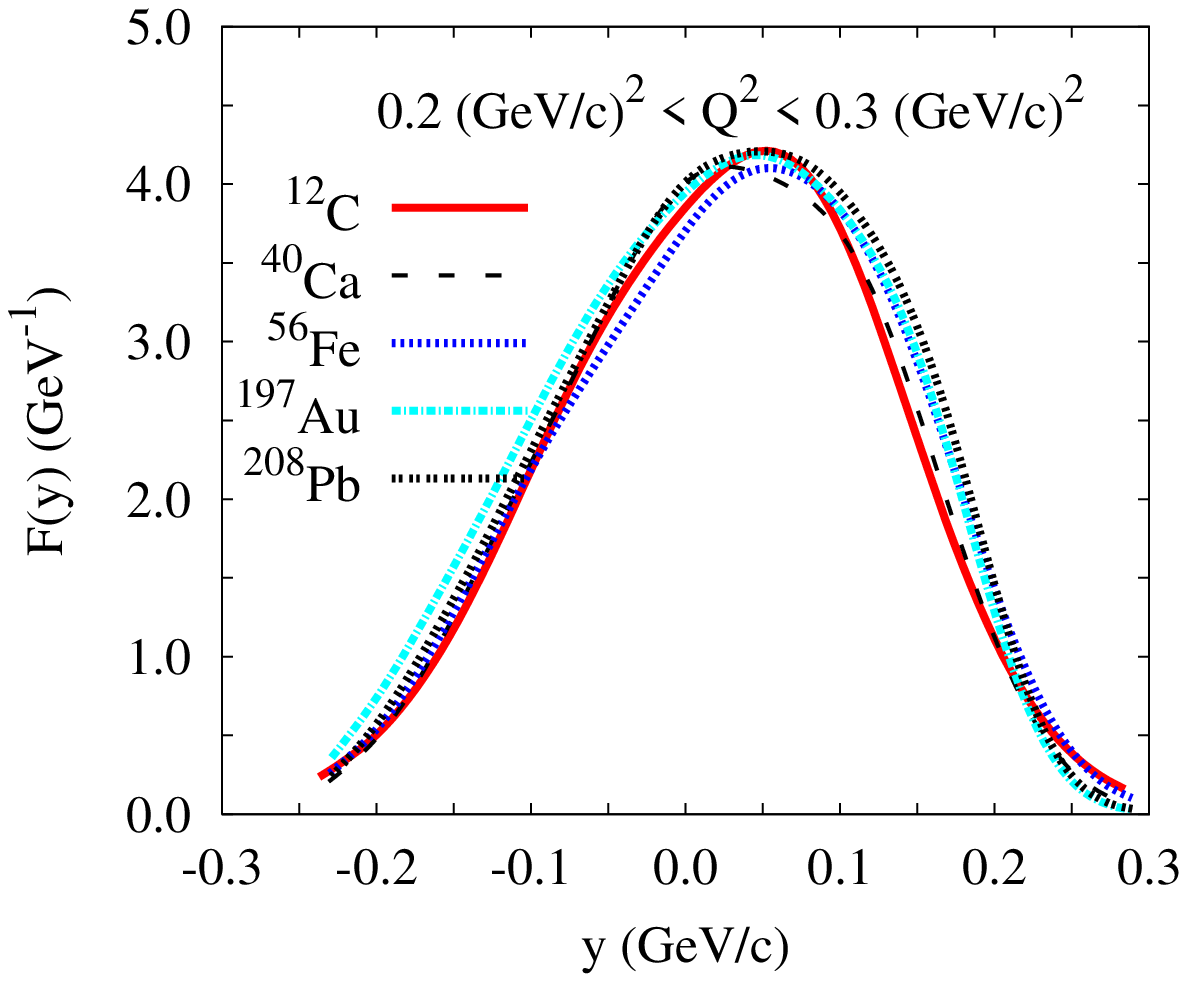}
\caption{The $y$-scaling functions for $^{12}$C, $^{40}$Ca,
$^{56}$Fe, $^{197}$Au, and $^{208}$Pb with incident electron
energy 750 MeV and scattering angle 45$^o$. The calculations do
not include the final state interaction of the outgoing nucleons
nor electron Coulomb distortion at the four momentum squared of
approximately 0.2 (GeV/c)$^2$ $< Q^2 <$ 0.3 (GeV/c)$^2$.}
\label{pi}
\end{figure}

In Fig. \ref{pi}, we calculate the $y$-scaling functions for
$^{12}$C, $^{40}$Ca, $^{56}$Fe, $^{197}$Au, and $^{208}$Pb by
neglecting both the final state interaction of the exiting
nucleons and the electron Coulomb distortion. This is normally
referred to as the plane wave impulse approximation (PWIA). The
squared four momentum transfer is approximately 0.2 (GeV/c)$^2$ $<
Q^2 <$ 0.3 (GeV/c)$^2$. The scaling function for all of these
cases as a function of the scaling variable $y$ are very
similar--with deviations from the mean less than 10\% which is
probably due to binding energy effects.   Note from the caption
that all of these examples were calculated with electron
scattering angles of $45^o$ since, as we will show later, there is
some breakdown of scaling at lower energies when the electron
scattering angle is changed significantly.

\begin{figure}
\includegraphics[width=1\linewidth]{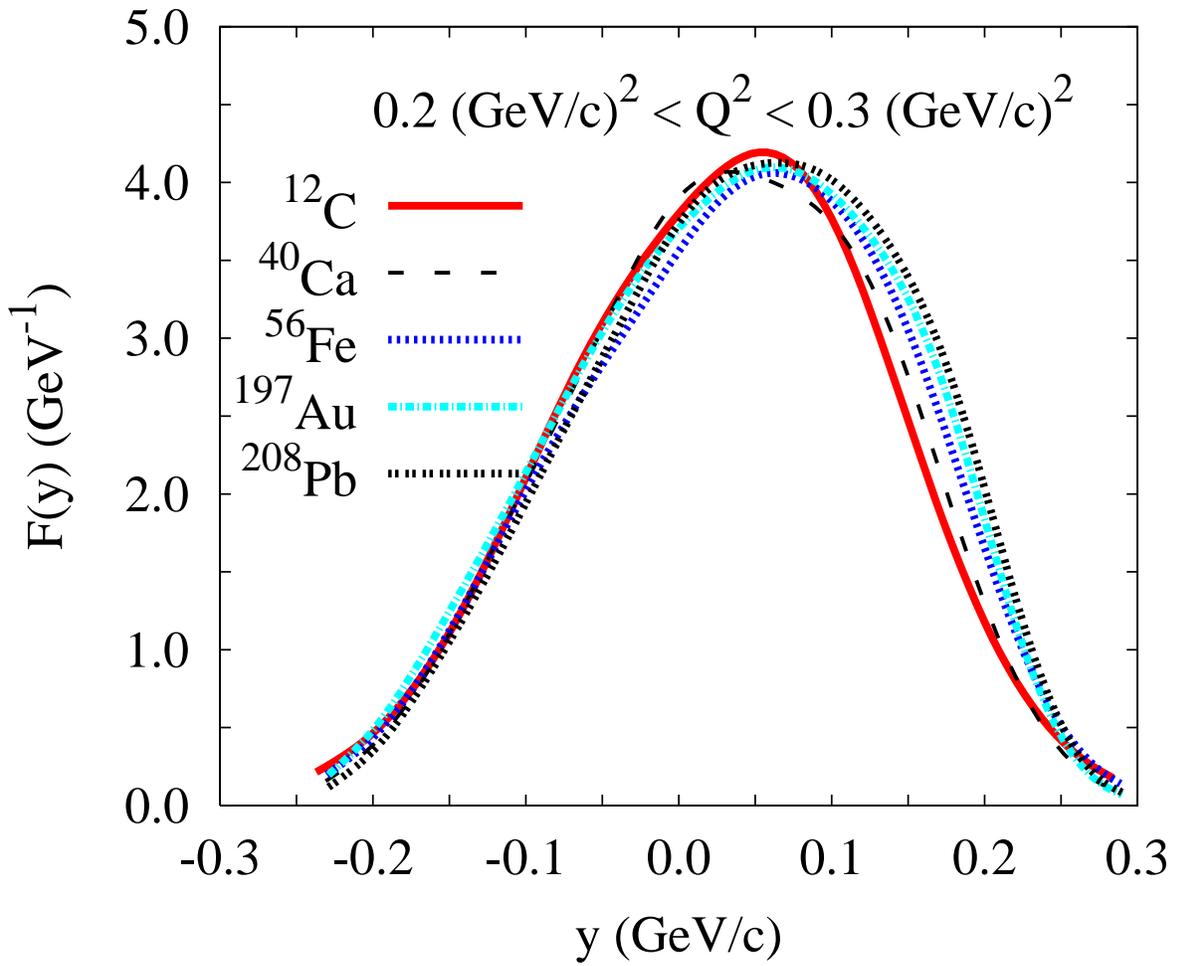}
\caption{The same as in Fig. \ref{pi} except electron Coulomb  distortion is included..} \label{di}
\end{figure}

\begin{figure}
\includegraphics[width=1\linewidth]{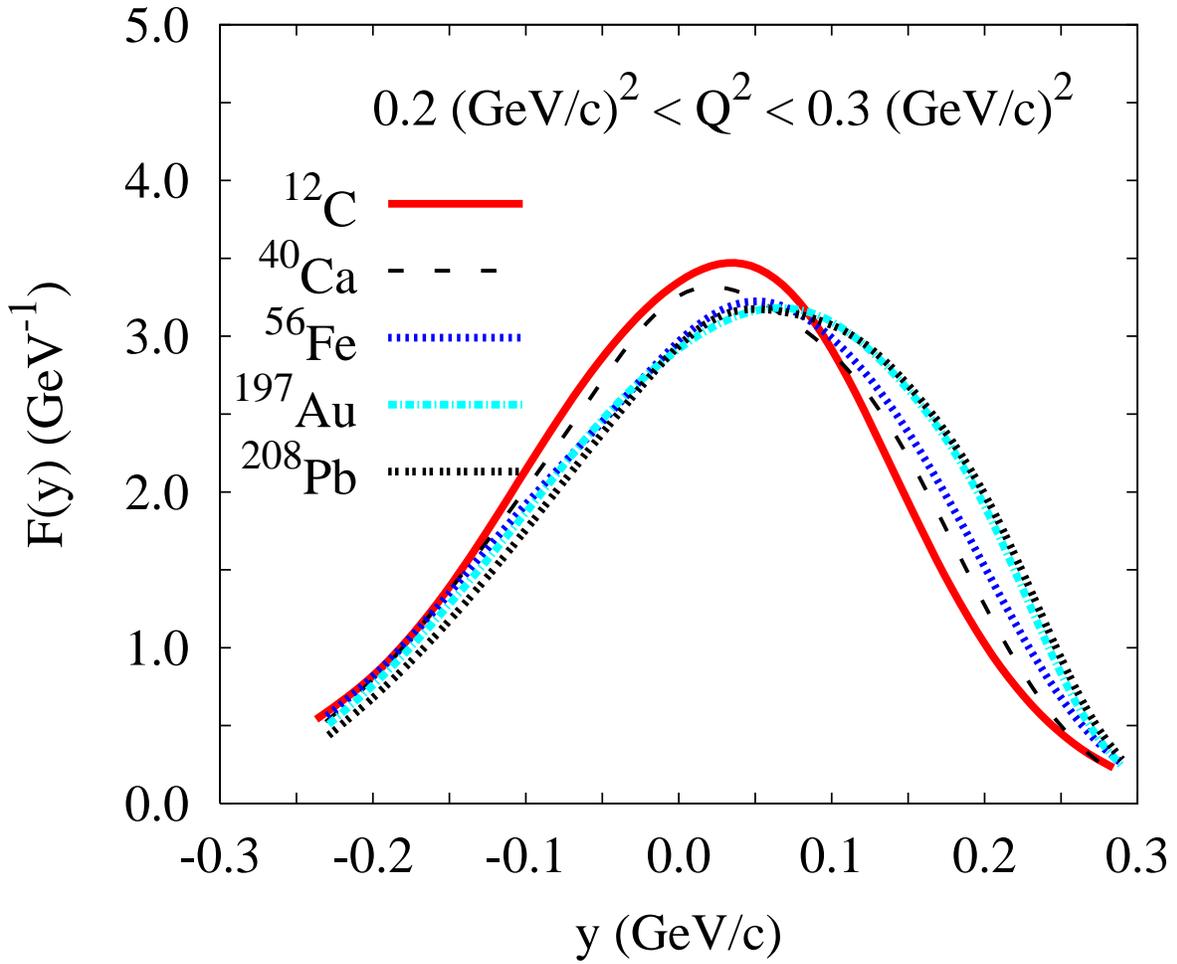}
\caption{The same as in Fig. \ref{pi} except  both electron
Coulomb distortion and the final state interaction of the outgoing
nucleons are included.} \label{dw}
\end{figure}

In Fig. \ref{di}, we show the same results  as in  Fig. \ref{pi},
but with electron Coulomb distortion turned on.  We refer to this
case as  the distorted wave impulse approximation (DWIA). For
heavier nuclei, electron Coulomb distortion shifts the scaling
curves toward the right side.  However, the deviations from
scaling due to Coulomb distortions do not seem to be too large and
the largest effect occurs for larger positive values of $y$. In
Fig. \ref{dw}, we show the same results as in Fig. \ref{pi}, but
with both the final state interaction and  the electron Coulomb
distortion included.

\begin{figure}
\includegraphics[width=1\linewidth]{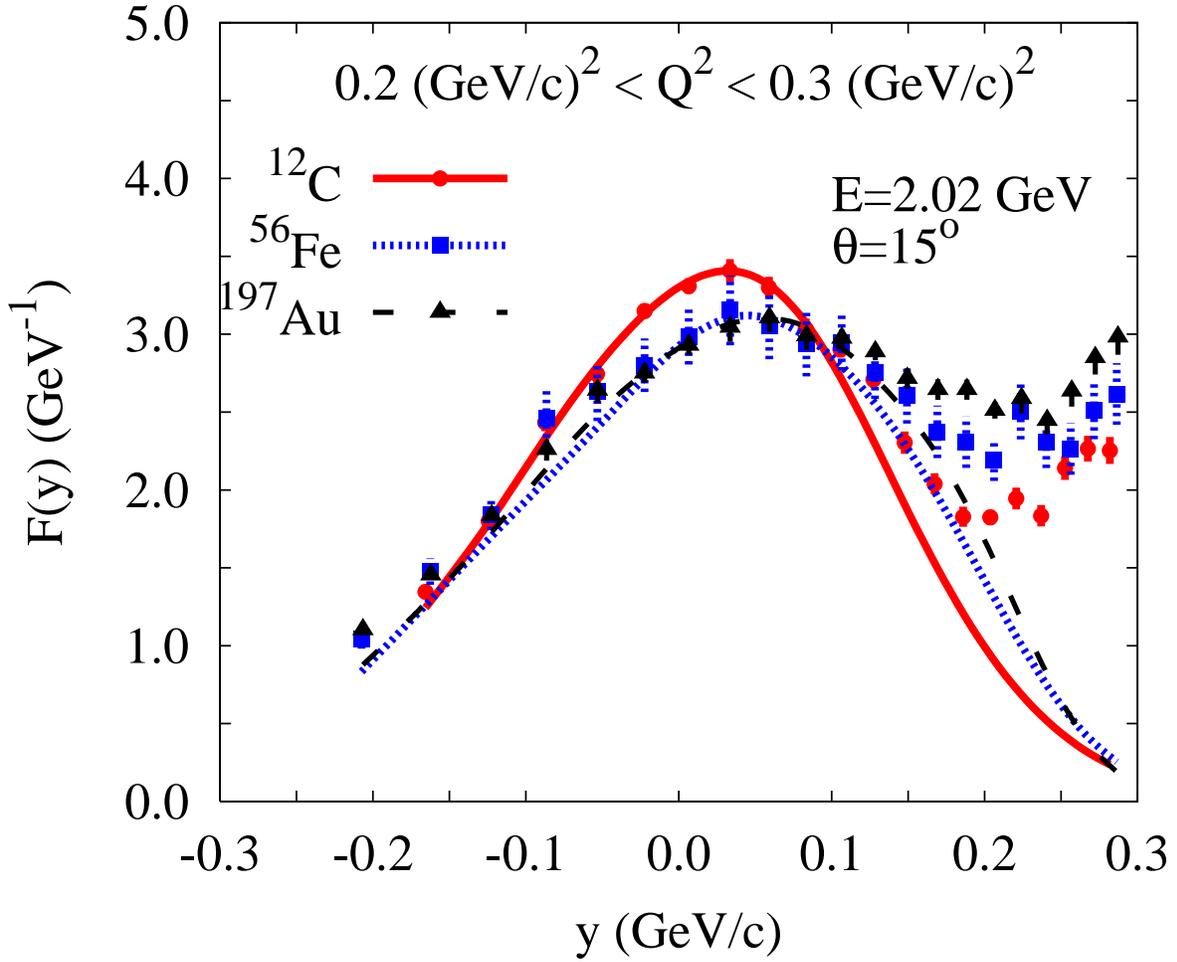}
\caption{The $y$-scaling functions for three different target
nuclei from $^{12}$C (sold and $\bullet$), $^{56}$Fe (dot and
$\blacksquare$), and $^{197}$Au (dash and $\blacktriangle$). The
electron incident energy is $E=2.02$ GeV and the scattering angle is
$\theta=15^{o}$. The experimental data are from SLAC \cite{day1}.}
\label{slac}
\end{figure}

Clearly the final state interaction in this kinematic
range leads to rather large violations of scaling which is also
observed experimentally. For example, Fig. \ref{slac} shows the
comparison of our theoretical results with the experimental data
for $^{12}$C, $^{56}$Fe, and $^{197}$Au measured at SLAC
\cite{day1}. The incident electron energy is $E=2.02$ GeV and the
electron scattering angle is $\theta=15^o$.  Under these kinematic
conditions, the energy transfer is below the pion production peak
(except for large positive values of $y$)and hence the
quasielastic peak is well separated from inelastic processes. Note
that our model, see also \cite{kim5,kim6}, describes the
quasielastic process for all three nuclei quite well and
furthermore, scaling is observed for the two heavier nuclei
($^{56}$Fe and $^{197}$Au) when the quasielastic process dominates
($y<0.1$).    However, the lighter nucleus ($^{12}$C ) does not
scale with the heavier ones.  This deviation is  due to a
combination of electron Coulomb distortion and final state
interaction effects.

\begin{figure}
\includegraphics[width=1\linewidth]{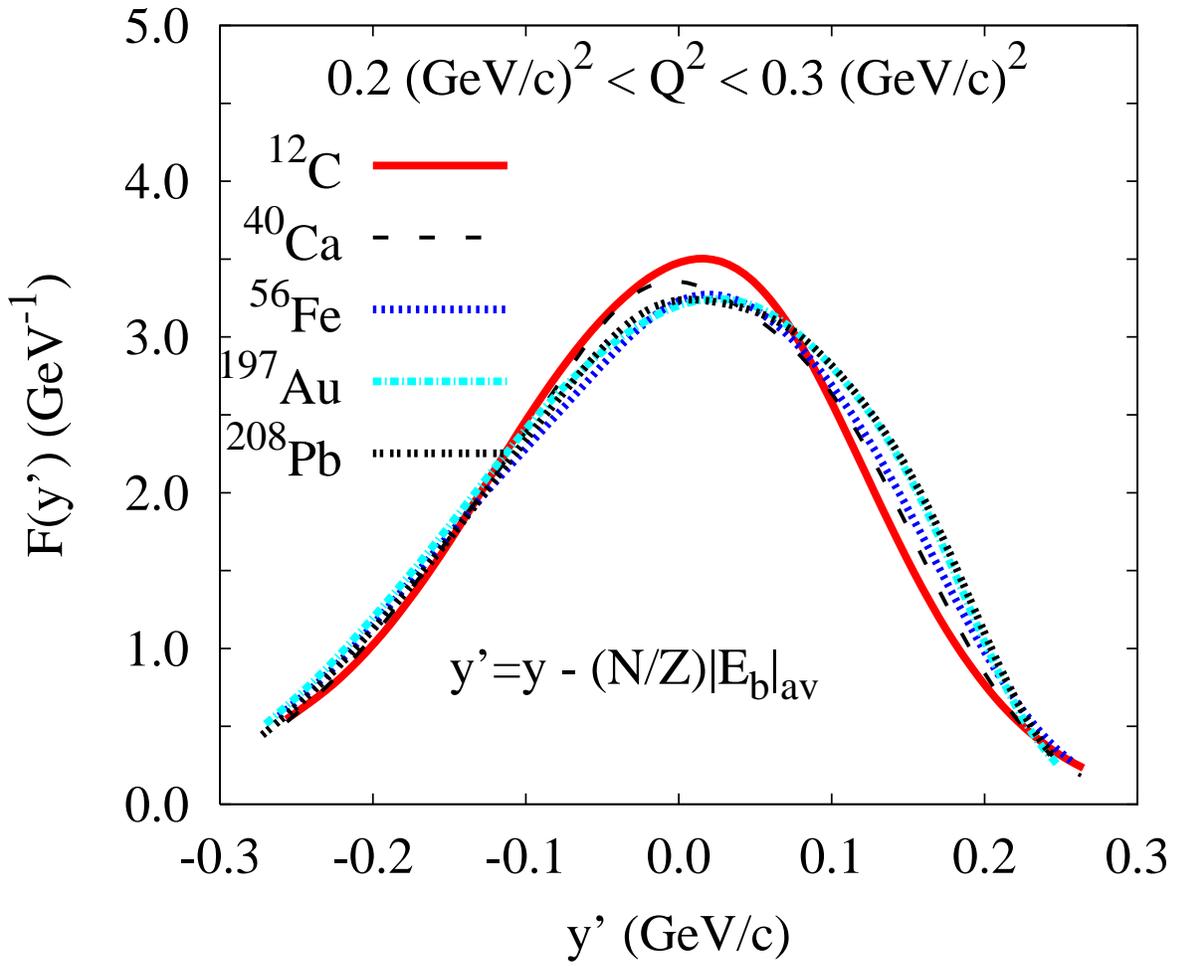}
\caption{The same kinematics as in Fig. \ref{dw} except plotted as
a function of the new $y'$-scaling variable.} \label{avnz750}
\end{figure}

In order to improve scaling we introduce a modified scaling
variable $y'$  given by $y'=y-(N/Z)|E_b|_{av}$ where $|E_b|_{av}$
is the average of the absolute binding energy for all the bound
nucleons. This approach is meant to remove differences among
nuclei including binding energy effects as suggested in Ref.
\cite{ciof2,ciof3} and may permit scaling for quasielastic
scattering from different nuclei. In Fig. \ref{avnz750}, we show
the new scaling functions for $^{208}$Pb, $^{197}$Au, $^{56}$Fe,
$^{40}$Ca, and $^{12}$C at the same kinematics as Fig. \ref{dw}.
Comparing Figs. \ref{dw} and \ref{avnz750} clearly shows that $y'$
improves scaling to some degree although some deviations from
scaling remain.

\begin{figure}[p]
\includegraphics[width=1\linewidth]{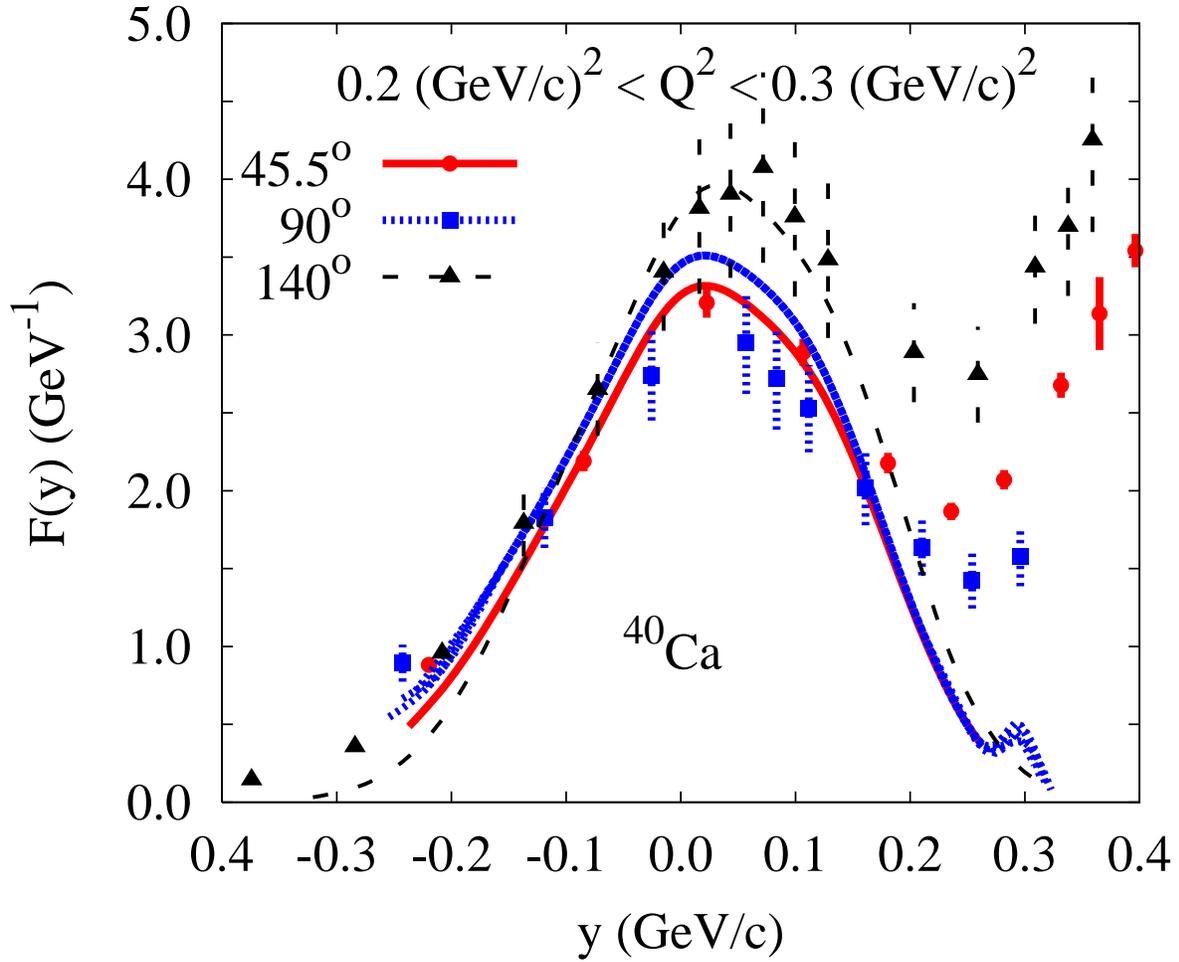}
\caption{The $y$-scaling functions for $^{40}$Ca at three
different electron energies, $E=739$ MeV and scattering angle
$\theta=45.5^{o}$ (solid and $\bullet$), $E=375$ MeV and
$\theta=90^o$ (dot and $\blacksquare$), and $E=367$ MeV and
$\theta=140^{o}$ (dash and $\blacktriangle$). The experimental
data are from Bates \cite{bates}.} \label{ca}
\end{figure}

\begin{figure}[p]
\includegraphics[width=1\linewidth]{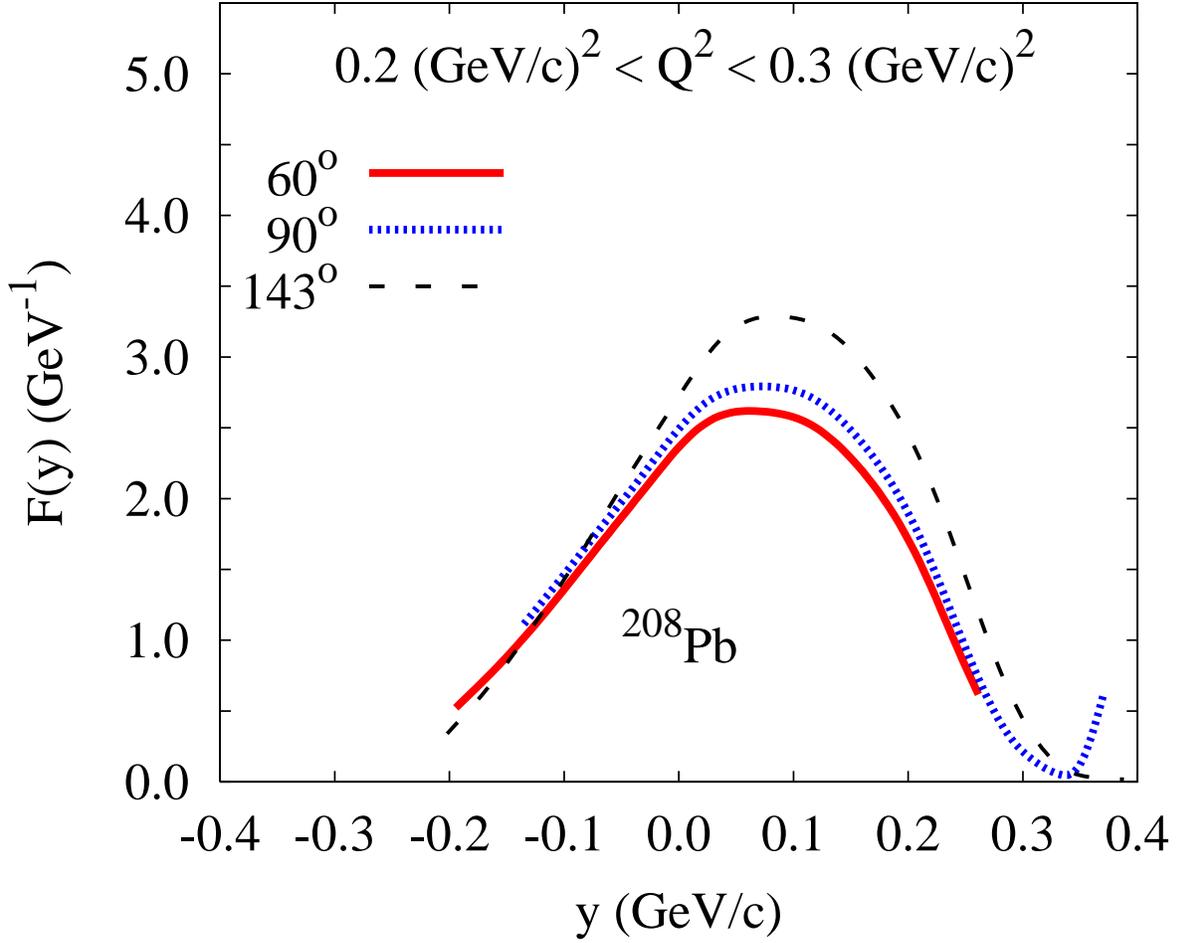}
\caption{The $y$-scaling functions for only the protons from
$^{208}$Pb at three different electron energies, $E=550$ MeV and
scattering angle $\theta=60^{o}$, $E=354$ MeV and $\theta=90^o$,
and $E=310$ MeV and $\theta=143^{o}$ with Coulomb distortion and final state
interaction included.} \label{pipb-pro}
\end{figure}

\begin{figure}[p]
\includegraphics[width=1\linewidth]{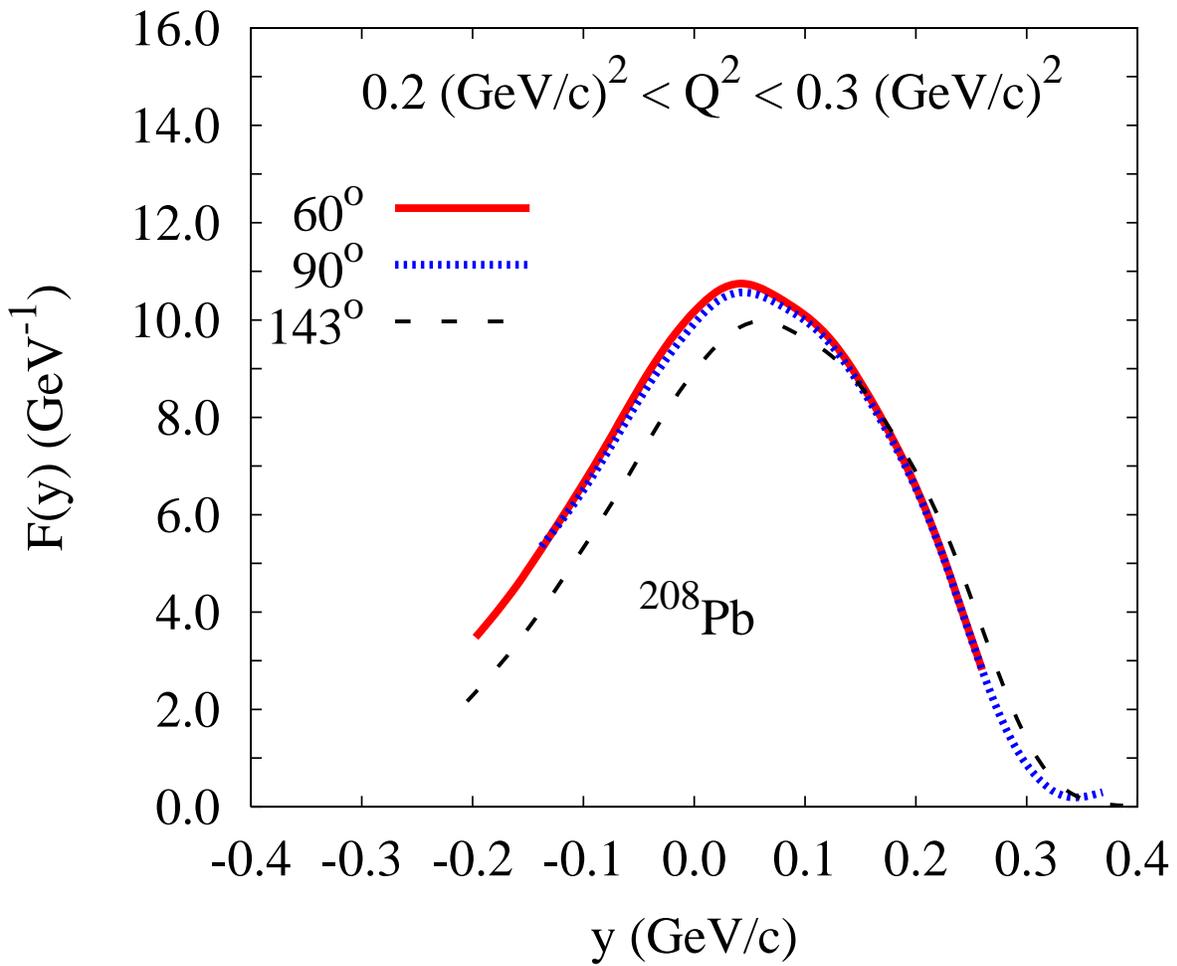}
\caption{The same as in Fig. \ref{pipb-pro} except for neutrons
only.} \label{pipb-neu}
\end{figure}

In Fig. \ref{ca}, we compare the scaling function to the
experimental data measured at Bates \cite{bates} on $^{40}$Ca for
three cases. The solid line and $\bullet$ are for incident
electron energy $E=739$ MeV and scattering angle $\theta=45.5^o$,
the dotted line and $\blacksquare$ for $E=372$ MeV and
$\theta=90^o$, and the dashed line and $\blacktriangle$ represent
$E=367$ MeV and $\theta=140^o$. The scaling functions have the
same shape and the peaks lie at the same position, but the
backward angle results (dashed curve and $\blacktriangle$)  have a
different magnitude from the  others. From these calculations, we
see that scaling at these somewhat lower energies is affected by
the electron scattering angle.  This result is in agreement with
the studies reported by Caballero et al. \cite{cabal} that final
state interactions affect longitudinal and transverse
contributions differentially since changing the electron
scattering angle changes the longitudinal and transverse
contributions to the cross section.  Note that these calculations
agree with the experimental data relatively well as in our
previous results \cite{bates,jin1,kim7}.  In order to investigate
the dependence of scaling at lower four momentum transfer values
on $\theta$, we calculated y-scaling with Coulomb distortion and
final state interactions
 for only the protons in $^{208}$Pb (Z=82 and
N=0)  as shown in Fig. \ref{pipb-pro} and for only neutrons ( Z=0
and N=126) as shown in Fig. \ref{pipb-neu}.   As expected, the
scaling function for the proton only results show a much larger
dependence on the scattering angle due to the changing
representation of the longitudinal and transverse response.  The
neutron only scaling function contains much less longitudinal
contributions (due to the motion of the magnetic moment) and
varies less with electron scattering angle.

\begin{figure}
\includegraphics[width=1\linewidth]{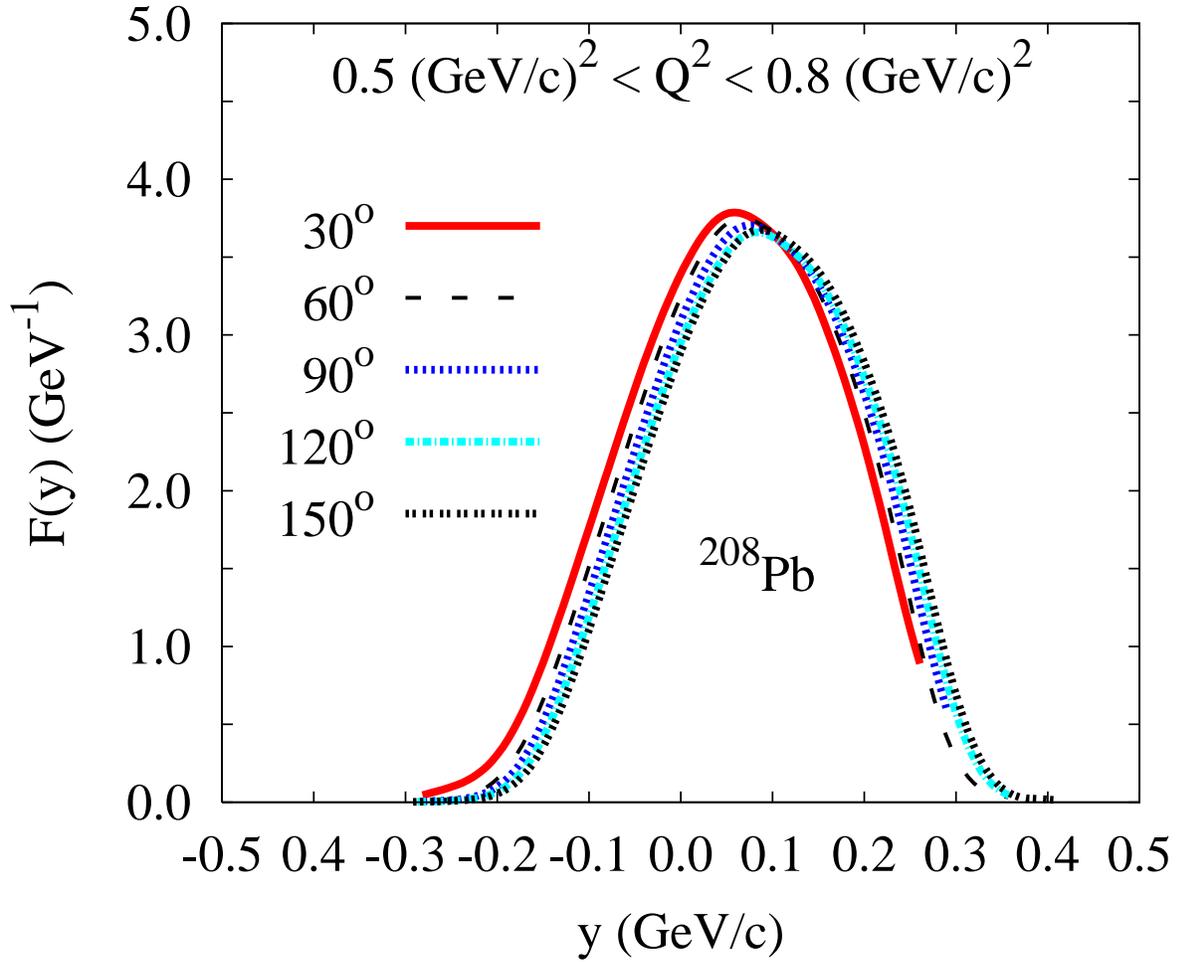}
\caption{The $y$-scaling functions for $^{208}$Pb both with
final state interaction and  Coulomb distortion at five
different electron kinematics, $E=1830$ MeV and scattering angle
$\theta=30^{o}$ (solid and red), 1000 MeV and $60^o$ (dash and
black), 740 MeV and $90^{o}$ (dot and blue), 630 MeV and $120^o$
(dash-dot and skyblue), and 580 MeV and $150^o$ (two-dot and
black). The range of the four momentum squared is approximately
0.5 (GeV/c)$^2$ $< Q^2 <$ 0.8 (GeV/c)$^2$.} \label{highpb}
\end{figure}

\begin{figure}
\includegraphics[width=1\linewidth]{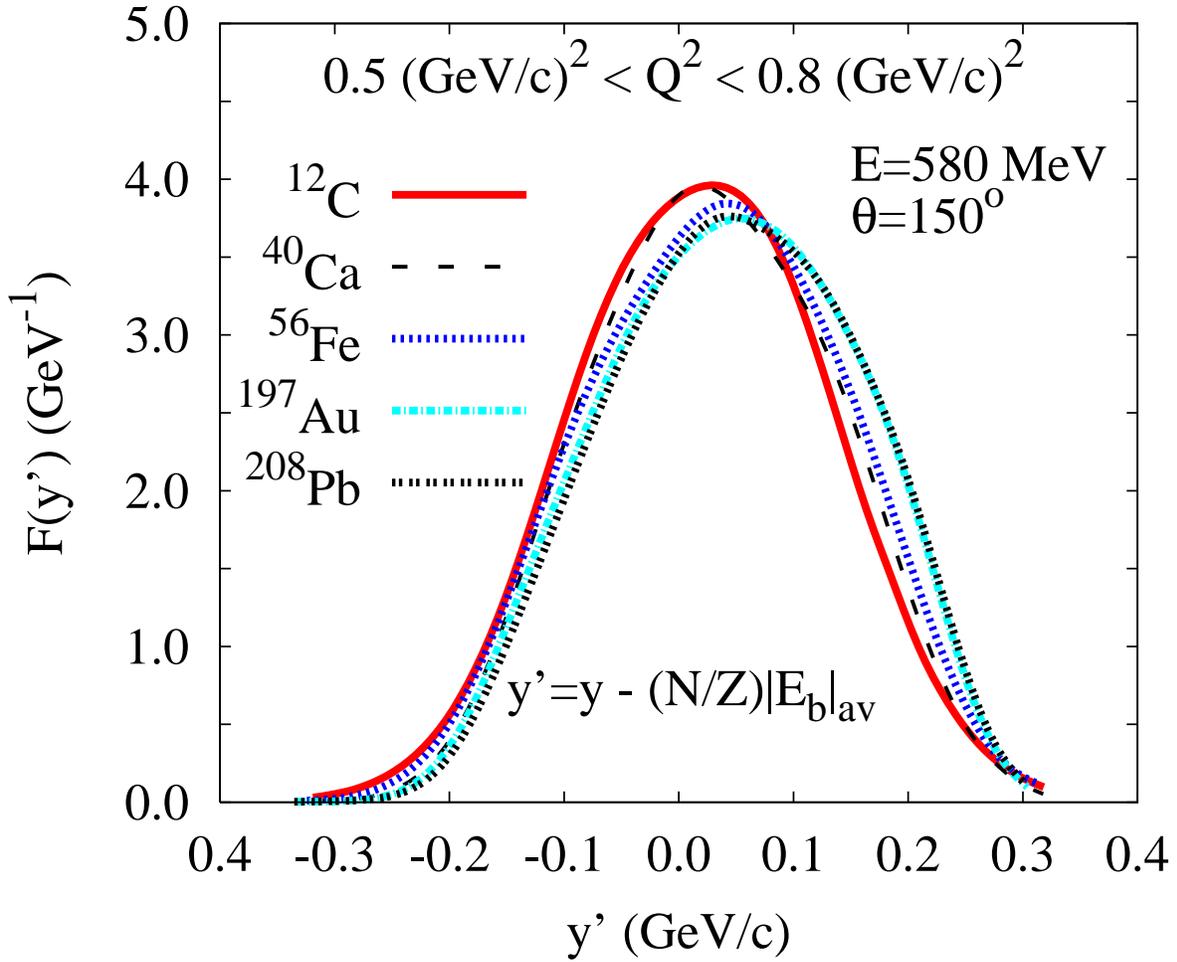}
\caption{The $y'$-scaling functions for several nuclei with
 final state interaction and  Coulomb distortion at the
incident electron energy $E=580$ MeV and scattering angle
$\theta=150^{o}$. The range of the four momentum squared is
approximately 0.5 (GeV/c)$^2$ $< Q^2 <$ 0.8 (GeV/c)$^2$.}
\label{avnz580}
\end{figure}

\begin{figure}
\includegraphics[width=1\linewidth]{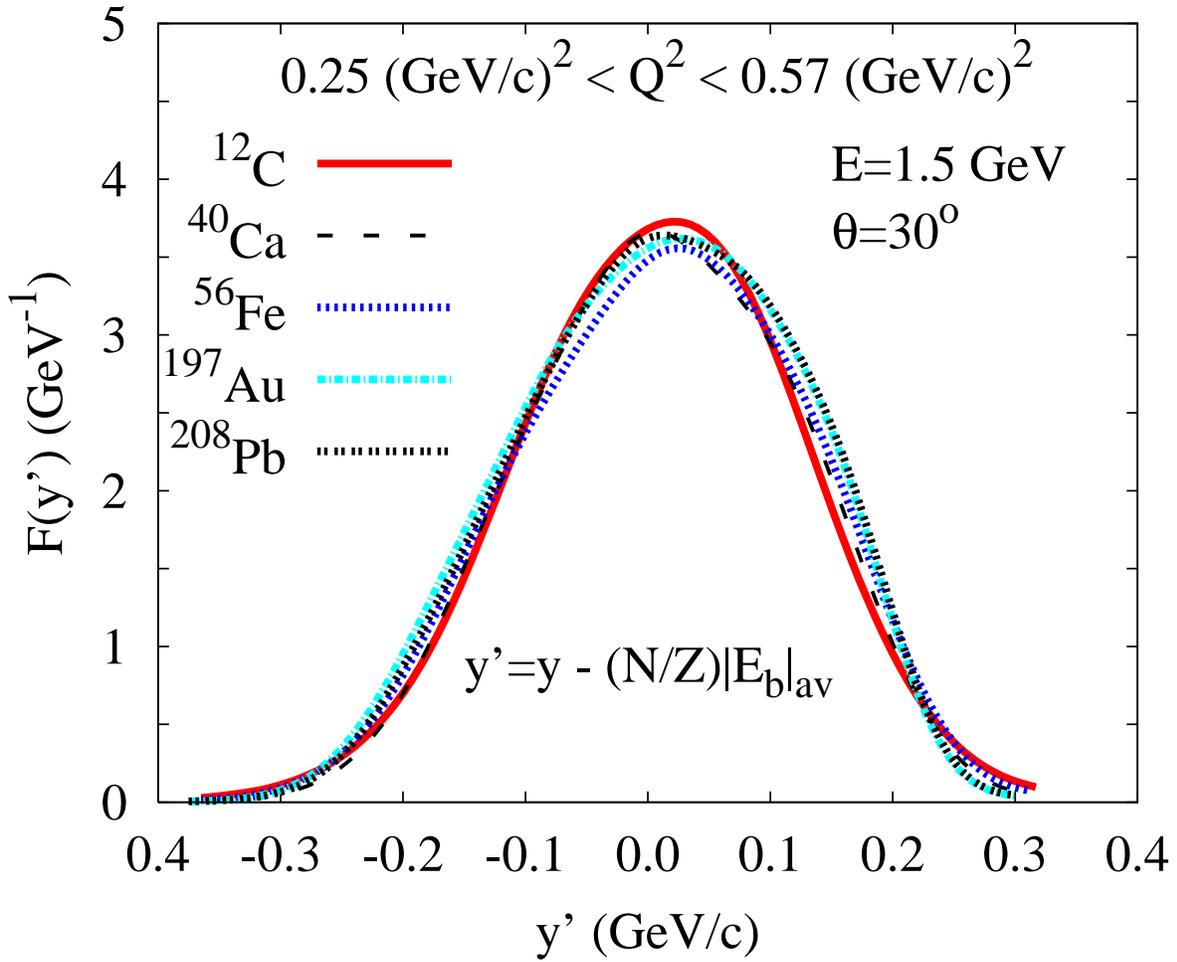}
\caption{The new $y'$-scaling functions  for the high electron
energy $E=1.5$ GeV and scattering angle $\theta=30^o$ from
several nuclei.} \label{avnzhigh}
\end{figure}

In  Fig. \ref{highpb}, we investigate scaling at larger values of
the four momentum transfer as a function of electron scattering
angle $\theta$ for quasielastic scattering on $^{208}$Pb when both
Coulomb distortion and final state interaction are included. Based
on this result and other cases we have examined, scaling is only
weakly dependent on $\theta$ at these higher kinematic values
although there is still some deviation at very large electron
scattering angles.  Furthermore, as shown in Fig.\ref{avnz580}
scaling for different nuclei still works well if all cases are
calculated at the same  backward scattering angle. Clearly as the
final state interaction decreases scaling improves. In Fig.
\ref{avnzhigh}, we show the new scaling functions at high incident
energy, $E=1.5$ GeV,  for the scattering angle $\theta=30^o$ in
terms of $y'$ as in  Fig. \ref{avnz750}. Clearly scaling is
greatly improved even at this intermediate four momentum transfer
region as long as we do not vary the electron scattering angles.

In summary, we have investigated $y$-scaling of the inclusive
$(e,e')$ reaction from $A$=12, 20, 56, 197, and 208 in the
quasielastic region which is kinematically isolated from inelastic
scattering.  We use a realistic nuclear model describes the
available data quite well and our calculations  include electron
Coulomb distortion.    As shown in Figs. \ref{ca} and
\ref{highpb}, the scaling function $F(y)$ is not independent of
electron scattering angle with particularly strong dependence at
backward scattering angles for lower values of the four momentum
transfer.  However, we have demonstrated that if one restricts the
variation of the electron scattering angles (thereby not changing
the longitudinal transverse mix significantly), scaling still holds
at these lower kinematic values.  Furthermore, as the energy
increases, the dependence of scaling on electron scattering angles
is greatly reduced.   We also found that $y$-scaling breaks down
at lower kinematic values due to the final state interaction.
Again, as the energy increases (and the final state interaction
weakens) the breakdown of scaling is greatly reduced.  We also
used our model to investigate the different binding energy effects
and final state interaction effects in different target nuclei.
We do find some improvement in scaling by modifying the definition
of the scaling variable $y$ to include binding effects.  Finally,
electron Coulomb distortion does disrupt scaling to some degree,
but does not play a significant role.

In conclusion, we confirm that the final state interaction is the
primary cause of the scaling violation in the quasielastic region
for lower energies, but we find that scaling improves dramatically
at larger four momentum transfer.  Our results show that at larger
four momentum transfer values, scaling holds across a range of
nuclei quite well.  Our model results,  which agree very well with the
available data,  suggest that scaling can be used at larger four
momentum transfer values to subtract out the quasielastic
contributions to the $(e,e')$ cross section from nuclei so that
inelastic contributions which contribute incoherently to the
measured cross section can be investigated.  Furthermore, our results
suggests that as the energies increase the scaling function obtained
from different nuclei and at different electron scattering energies differ
at most by only 10 to 20 \%.

\section*{Acknowledgements}
One of us (Wright) wishes to acknowledge partial support of this
work from a research grant from the Department of Energy.

\end{document}